\documentclass[anonymous,review,conference]{IEEEtran}
\IEEEoverridecommandlockouts
\usepackage{cite}
\usepackage{amsmath,amssymb,amsfonts}
\usepackage{algorithmic}
\usepackage{graphicx}
\usepackage{textcomp}
\usepackage{xcolor}

\usepackage{CJKutf8}
\usepackage{makecell}
\usepackage{subcaption}
\usepackage{xspace}
\usepackage{float}
\usepackage[hidelinks]{hyperref}

\usepackage{tipa}

\usepackage{multirow}
\usepackage{multicol}
\usepackage{enumitem}
\usepackage{url}
\usepackage{colortbl}
\usepackage{tcolorbox}
\usepackage{booktabs}
\usepackage{balance}

\usepackage[T1]{fontenc}

\def\BibTeX{{\rm B\kern-.05em{\sc i\kern-.025em b}\kern-.08em
    T\kern-.1667em\lower.7ex\hbox{E}\kern-.125emX}}
\begin{document}

\title{Metamorphic Testing for Audio Content Moderation Software}

\author{\IEEEauthorblockN{Wenxuan Wang$^{1}$, Yongjiang Wu$^{2}$, Junyuan Zhang$^{2}$, Shuqing Li$^{2}$, Yun Peng$^{2}$\\ Wenting Chen$^{3}$, Shuai Wang$^{4}$, Michael R. Lyu$^{2}$}

\IEEEauthorblockA{$^1$ School of Information, Renmin University of China, China}

\IEEEauthorblockA{$^2$ Department of Computer Science and Engineering, The Chinese University of Hong Kong, China}

\IEEEauthorblockA{$^3$ Department of Electrical Engineering, City University of Hong Kong, China}

\IEEEauthorblockA{$^4$ Department of Computer Science and Engineering, Hong Kong University of Science and Technology, China}

\IEEEauthorblockA{wangwenxuan@ruc.edu.cn, \{yojwu,1155191545\}@link.cuhk.edu.hk, shuaiw@cse.ust.hk, \{sqli21,ypeng,lyu\}@cse.cuhk.edu.hk}

}

\maketitle

\newcommand{\methodname}{MTAM\xspace}

\begin{abstract}
The rapid growth of audio-centric platforms and applications such as Whatsapp and Twitter has transformed the way people communicate and share audio content in modern society.
However, these platforms are increasingly misused to disseminate harmful audio content, such as hate speech, deceptive advertisements, and explicit material, which can have significant negative consequences (e.g., detrimental effects on mental health).
In response, researchers and practitioners have been actively developing and deploying audio content moderation tools to tackle this issue. 
Despite these efforts, malicious actors can bypass moderation systems by making subtle alterations to audio content, such as modifying pitch or inserting noise. 
Moreover, the effectiveness of modern audio moderation tools against such adversarial inputs remains insufficiently studied.
To address these challenges, we propose \textit{\methodname}, a \underline{M}etamorphic \underline{T}esting framework for \underline{A}udio content \underline{M}oderation software.
Specifically, we conduct a pilot study on $2000$ audio clips and define 14 metamorphic relations across two perturbation categories: Audio Features-Based and Heuristic perturbations. 
\methodname applies these metamorphic relations to toxic audio content to generate test cases that remain harmful while being more likely to evade detection.
In our evaluation, we employ \methodname to test five commercial textual content moderation software and an academic model against three kinds of toxic content.
The results show that \methodname achieves up to $38.6\%$, $18.3\%$, $35.1\%$, $16.7\%$, and $51.1\%$ error finding rates (EFR) when testing commercial moderation software provided by Gladia, Assembly AI, Baidu, Nextdata, and Tencent respectively, and it obtains up to $45.7\%$ EFR when testing the state-of-the-art algorithms from the academy.
In addition, we leverage the test cases generated by \methodname to retrain the model we explored, which largely improves model robustness (nearly $0\%$ EFR) while maintaining the accuracy on the original test set.
We release the code and experiment data to facilitate future research\footnote{https://github.com/yjwu04/MTAM}.
\end{abstract}


\section{Introduction}
\label{sec-introduction}
Recently, the digital world has been rapidly evolving, with social media platform users and electronic gamers increasing tremendously, greatly enhancing modern audio or video communication worldwide.
For example, the monthly active users on Tiktok has grown from $55$ million in $2018$ to $1.04$ billion in $2024$ ~\cite{tiktok2024}.
Additionally, the advancement of text-to-speech (TTS) technology has made audio creation much easier ~\cite{Ren2020FastSpeech2F}.
For example, around $30K$ words are processed and transformed into audio on a website\footnote{http://www.ttsonline.cn/} per day.
However, the online TTS tools are free to use without censorship.
As a result, the toxic speech prevalence among these social media platforms or community forums intensifies under this trend due to the anonymity of the web.
Audio toxic contents typically refer to three major kinds of audios: (1) \textit{abusive language and insult speech}, which are offensive audios with aggressive tone targeting specific individuals, such as politicians, celebrities, religions, nations, and the LGBTIQA+ \cite{Badjatiya2017DeepLF}; (2) \textit{pornography}, which is often sexually explicit, associative, and aroused such as panting~\cite{Rowley2006LargeSI}; and (3) \textit{malicious advertisement}, which are voice segments sent on social media, video clips inserted with fake advertisements, and suspicious phone calls with illegal purposes, such as phishing and scam links, pyramid selling, false advertising and illegal information dissemination~\cite{Li2012KnowingYE}.
These toxic contents can lead to highly negative impacts.
Specifically, Munro~\cite{children2011} studied the ill effects of online \textit{insult speech} on children and found that children may develop depression, anxiety, and other mental health problems.
\textit{Pornography} can cause significant undesirable effects on the physical and psychological health of children \cite{Yu2016InternetMI}.
\textit{Malicious advertisements} remain a notorious global burden, accounting for up to $85\%$ of daily message traffic \cite{spam2022}.
Moreover, these toxic contents have even increased the number of criminal cases to a certain extent~\cite{Chen2020AutomaticDO}.
These studies show that harmful audio poses significant risks to social harmony. As a result, content moderation tools designed to identify and block such material have garnered significant attention from both academia and industry.

Typical audio moderation software first detects toxic content and then blocks it or warns the users before showing it.
Toxic content detection, central to audio moderation, is often approached as a classification problem. Various deep learning models, including convolutional neural networks, Long-Short-Term-Memory (LSTM) models, and Transformers, associated with Automatic Speech Recognition (ASR) have been used to address it~\cite{Mishra2019TacklingOA, Schmidt2017ASO, Wu2018TwitterSD, Gupta2022ADIMAAD}.
Recently, the development of ASR and pre-trained language models (e.g., BERT~\cite{Devlin2019BERTPO} and RoBERTa~\cite{Liu2019RoBERTaAR}) has significantly improved the held-out accuracy of toxic content detection.
Due to recent advancements, companies have also widely implemented commercial content moderation software in their products, such as Baidu~\cite{notrobustbaidu}, Tencent~\cite{tenaudmod2025}, and Gladia~\cite{glaAudmod2025}.

However, the mainstream content moderation software is not robust enough~\cite{notrobustbaidu}.
For example, Baidu content moderation software cannot understand many languages, leaving non-Chinese speaking users more susceptible to harmful posts \cite{notrobustbaidu}.
In addition, toxic audio can bypass mainstream content moderation software by applying simple homophonous transformations. For example, changing ``bitch'' to ``beach'' in English or increasing pitch on toxic words through a simple voice changer during live streaming~\cite{bypassMod2022}.
The vital initial step is creating a testing framework for content moderation software, much like traditional software, to tackle this issue.

A lack of testing frameworks for audio content moderation software persists, partly due to the complexity of the problem.
First, most audio moderation systems are based on ASR for the first step, but ASR itself nowadays could make mistakes, identifying a secure audio segment as toxic, even with sufficient fine-tuning ~\cite{Radford2022RobustSR}.
In addition, a recent study~\cite{AEON2022ISSTA} reported that $44\%$ of the test cases generated by the State-of-the-Art (SOTA) approaches of automated testing techniques for Natural Language Processing (NLP) softwares are false alarms, which are cases often have inconsistent semantics or incorrect grammar, making these approaches less effective.
Moreover, most of the testings are done under the assumption of a normal environment with pure voices or slight perturbations~\cite{Wang2023SoftwareTW}.
However, the actual application scenario could be more extreme than the test environment, especially when encountering adversarial attacks to bypass the moderation.

In this paper, we propose \textit{\methodname}, an \underline{M}etamorphic \underline{T}esting framework for \underline{A}udio content \underline{M}oderation software.
In particular, in order to develop a comprehensive testing framework for audio moderation software, we should begin with an understanding of what kinds of transformations real users might apply to bypass moderation.
Therefore, we examined 2,000 intentionally perturbed audio clips, designed to bypass moderation, collected from real users in a pilot study (Section \ref{sec-mrs}) and summarise $14$ deformation relations at two perturbation levels, audio-feature-based level and text-based level, so that the method can provide metamorphic relations reflecting real-world user behaviours and designed for audio moderation software.
\methodname employs these deformation relations on toxic contents to generate test cases that are still toxic (i.e., being easily identifiable by humans) but with the potential to bypass moderation. 
We implement the metamorphic relations for two languages, English and Chinese, because English is a representative language based on the alphabet, while Chinese is a representative language based on the pictograph. 
For audio content moderation in Chinese, we apply \methodname to test three commercial audio content moderation software against three typical kinds of toxic content (i.e., abusive language, malicious advertisement, and pornography).
And for audio content moderation in English, we test two software and one audio moderation model, replacing malicious advertisements with socially sensitive issues as the tested category from the listed three kinds of toxic content.
The results show that \methodname achieves up to $38.6\%$, $18.3\%$, $35.1\%$, $16.7\%$, and $51.1\%$ error finding rates (EFR) when testing commercial moderation software provided by Gladia, Assembly AI, Baidu, Nextdata, and Tencent respectively.
Additionally, we also utilize the data generated by \methodname to retrain an audio moderation model. The result shows that the model robustness achieves significant improvement with steady accuracy on the initial test set.

The main contributions of this paper are as follows:
\begin{itemize}[leftmargin=*]
    \item The introduction of the first comprehensive testing framework, \methodname, for audio content moderation software validation.
    \item A pilot study was conducted on 2,000 audio clips, sourced either from real-world recordings or from daily text messages converted via text-to-speech (TTS), that were intentionally designed to bypass moderation. This study led to the identification of fourteen metamorphic relations, which in turn guided the implementation of \methodname in two languages: English and Chinese.
    \item A broad assessment of \methodname on five commercial content moderation software and an academic audio moderation model, indicating that toxic audios generated by \methodname can easily bypass moderation and improve the robustness of the audio moderation models.
\end{itemize}

\noindent \textbf{Content Warning}: This article presents examples of offensive, abusive or pornographic expressions, for which we apologise. These examples are quoted verbatim. In addition, in order to conduct this study safely, we took the following precautions with the participants: (1) at each stage, we prompted the researchers and annotators with content warnings and told them that they could leave at any time during the study; and (2) at the end of the study, we provided them with counselling to alleviate their mental stress.
\section{Challenges to Reliability}
\label{sec-discuss}


Although MTAM addresses the lack of testing frameworks, its reliability may face several challenges.
\textbf{The initial challenge} is that the test cases produced by \methodname, after undergoing numerous modifications, might turn out to be "non-toxic," which could result in false positive outcomes.
Moreover, these cases may be classified as another type of toxic message without much semantic meanings.
Even worse, some of our metamorphic relations (MRs) could mislead the software to identify a non-toxic message as toxic as well.
These actually relates to adversarial attack field in audio, which is supplementary to our work.
To address this concern, we carried out a user study to confirm whether the generated test cases are indeed toxic and avoid adversarial attack examples.
We also requested the annotators to indicate whether the test cases represent inputs from actual users.
The findings indicate that the generated test cases are both toxic and realistic.
\textbf{The second challenge} is that we have implemented \methodname for two specific languages, which may not be applicable to other natural languages.
To mitigate this issue, we carefully selected these two natural languages which are the largest language families in the world with one represents an alphabet-based system while the other is a pictograph-based system.
Furthermore, we are confident that our MRs can apply to other languages since languages exhibit similar characteristics (e.g. homophone, environmental noises, four basic features).
\textbf{The third threat} lies in our assessment of five audio moderation systems, which may not accurately reflect \methodname's effectiveness on different systems.
To address this concern, we evaluate both commercial content moderation software and a SOTA academic model.
In particular, we test content moderation software provided by three big companies, which already have widely applications to defend malicious inputs.
A future direction of our work could be to test more commercial software and research models to further mitigate this threat.
\textbf{The fourth concern} is that our \methodname could be outdated with the bypass techniques evolving.
However, our work can be abstracted and generalized in a comprehensive workflow: Analyze user behaviors, compile and design the MRs, create test cases, and utilize failure scenarios to enhance robustness.
This workflow can be adapted in latest bypass techniques proposed to design new MRs.

Another promising and valuable avenue to pursue is the automated MR generation.
Research listed below mainly focuses on automated generation of a specific kind of MRs (e.g., polynomial MRs~\cite{Zhang2014SearchbasedIO, Zhang2019AutomaticDA} or automated MR generation leveraging software redundancy~\cite{Carzaniga2014CrosscheckingOF}. Given the various challenges associated with automated MR generation for content moderation software, we consider it a significant area for future research.



\section{Background}
\label{sec-backgound}

\subsection{Audio Moderation}

\subsubsection{Automatic Speech Recognition}

Automatic Speech Recognition (ASR) is a technology that converts spoken language into text.
Traditional ASR models utilize Mel-Scale Frequency Cepstral Coefficients
(MFCC) feature extraction, an acoustic model such as Gaussian Mixture Models-Hidden Markov Models (GMM-HMM), and a language model such as n-grams~\cite{Malik2020AutomaticSR}.
First of all, the sound is collected and represented as a sonogram.
Then, after the Fourier Transform, the sonogram is converted to a spectrum.
A spectrogram is finally used to represent the voice segment by concatenating all spectra.
MFCC feature extraction transforms the sound into high-dimensional vectors~\cite{ittichaichareon2012speech}.
GMM represents the probability distribution of acoustic features by modeling the variations in speech sounds using a mixture of Gaussian distributions, while HMM captures the temporal sequence of speech sounds by simulating the progression of speech through different states, with each representing a phoneme or part of a phoneme~\cite{swietojanski2013revisiting}.
After these vectors are converted into a sequence of phonetic units, the n-grams language model analyzes them and predicts the most likely texts~\cite{Malik2020AutomaticSR}. 
However, with the development of deep learning, neural networks replace MFCC feature extraction, acoustic models, and language models to form an end-to-end speech-to-text model such as the RNN-T model~\cite{jain2020contextual}.

\subsubsection{Audio Moderation Software}

Major corporations like Google, Baidu, and Tencent have integrated commercial audio moderation tools into their products.
As described in their technical documents, these tools typically use a hybrid classification system combining neural networks and rule-based algorithms, harnessing the strengths of each~\cite{GoogleMod2023,BaiduMod2024,TencentMod2024}.
Neural networks excel at grasping context and semantics, while rule-based methods efficiently implement specific user requirements. 
For instance, Baidu’s content moderation relies on deep learning models alongside an extensive list of prohibited terms~\cite{BaiduMod2024}.
However, the software's specific implementation architecture and algorithms have not been published. 
Nevertheless, the software's API call processes and feedback results suggest that most of them utilize ASR first to transform the audio into texts and then perform textual content moderation instead of operating the audio features directly.
After ASR, there are generally two categories of models for textual content moderation: \textit{feature engineering-based} models and \textit{neural network-based} models.

\noindent \textbf{Feature Engineering-Based Models}. Feature engineering-based models train their toxic content classification models based on manually constructed features.
Specifically, textual feature engineering can be further divided into \textit{rule-based} methods and \textit{statistical} methods.
The core of rule-based methods is pre-defined rules or dictionaries of banned words, while statistical methods leverage different statistics of the textual data.

\noindent \textbf{Neural Network-Based Models}. Advancements in text representation learning have spurred researchers to explore neural network-based models for textual content moderation~\cite{Djuric2015HateSD,Badjatiya2017DeepLF,pennington2014glove} 
With the help of the pre-trained language models (e.g., BERT \cite{Devlin2019BERTPO} and RoBERTa \cite{Liu2019RoBERTaAR}), researchers fine-tune these models on a downstream dataset and achieved remarkable performance on textual content moderation tasks.

\subsection{Metamorphic Testing}


Metamorphic testing \cite{Chen2020MetamorphicTA} is a property-based testing technique, which is an effective testing approach widely applied to solve the oracle problem.
It focuses on detecting violations of \textit{metamorphic relations} (MRs) across multiple runs of the specific software being measured.
Concretely, MR depicts the relationship between input-output pairs of software, which means, given a test case, metamorphic testing transforms it into a new test case via a pre-defined transformation rule and then checks whether the corresponding outputs of these test cases returned by the software exhibit the expected relationship.

In recent years, metamorphic testing has already been employed to validate Artificial Intelligence (AI) software.
Studies in this area mainly center on aiming to automatically report erroneous results returned by AI software through developing novel types of MR.
In concrete terms, Santos et al. \cite{10.1145/3425174.3425226} choose metaphoric testing to verify a machine learning breast cancer diagnostic application. 
Deng et al. \cite{9477683} designed and evaluated a metamorphic testing framework to express domain-specific constraints and found violations of such constraints in order to improve the self-driving system.
Jiang, Tsong and Wang \cite{JIANG2022106966} incorporated metamorphic testing into the testing sentiment analysis systems, and observed a significant effectiveness.
Yuan, Pang and Wang \cite{10.1145/3551349.3561157} managed to unveil hidden Deep Nerual Network (DNN) defects with decision-based metamorphic testing. 
Dwarakanath et al. \cite{Dwarakanath2018IdentifyingIB} proposed eight MRs to test SVM-based and ResNet-based image classifiers. 


\section{\methodname}
\label{sec-mrs}

This section first introduces a pilot study on audio clips collected from real users (Section \ref{sec:pilotstudy}).
Then we introduce fourteen metamorphic relations (MRs) that are inspired by the pilot study and previous studies~\cite{peterson2015metamorphic,Moreira2020TestingAS,Wang2022SRMTAM,Tan2022AdversarialAA,Mauch2013TheAD, Parascandolo2016RecurrentNN,Salamon2016DeepCN,Wei2020ACO}.
These MRs can be grouped into three categories according to the perturbation level performed: (1) Basic Signal-Level Perturbations (Section \ref{sec:basis}), (2) Compound Signal-Level Perturbations (Section \ref{sec:compound}), and (3) Linguistic-Form Perturbations (Section \ref{sec:heuristic}).

\subsection{Pilot Study}
\label{sec:pilotstudy}

\begin{table}
\caption{Overview of the perturbation taxonomy under MRs.}
\label{tab:mrs}
\centering
\begin{tabular}{l l}
\toprule
\bf Perturbation Type & \bf Perturbation Method\\
\midrule
\multirow{5}{*}{Basic Signal-Level} & Time-Domain Perturbation \\ 
& Space-Domain Perturbation \\ 
& Frequency-Domain Perturbation \\ 
& Injection \\ 
& Amplitude Adjustment \\ 
\midrule
\multirow{7}{*}{Compound Signal-Level} & Compression \\ 
& Ring Modulation \\ 
& Base Boost \\ 
& Tremolo \\ 
& Distortion \\ 
& Echo \\ 
& Reverb \\ 
\midrule
\multirow{2}{*}{Linguistic-Form}
& Homophone Substitution \\
& Benign Discontinuity \\

\bottomrule
\end{tabular}
\end{table}

In this work, an ideal metamorphic relation should satisfy that the original audio and the perturbed audio should be classified as the identical label by annotators, but different labels by the audio moderation software. So, the MRs in the experiment should follow the following criteria:    
\begin{itemize}[leftmargin=*]
    \item \textit{Semantically equivalent}: the perturbed test cases should possess
    the identical semantic meaning as the original audio.
    \item \textit{Realistic}: the perturbed test cases should reflect the possible inputs by real users, and the corresponding MR should be derived from the audio moderation bypass methods or the audio processing approaches in real life.
    \item \textit{Unambiguous}: the MRs in the perturbed test cases should be defined clearly and precisely.
\end{itemize} 

For the purpose of designing expected MRs in the perturbation process, we initially conducted a pilot study to simulate real-world scenarios. The study is intended to investigate what kind of perturbations users may employ to allow the toxic audio to bypass the audio moderation software. The audios for the pilot study sources are sourced from a large number of real users on some social media platforms and several video websites:
\begin{itemize}[leftmargin=*]
    \item Youtube\footnote{https://www.youtube.com/} is currently the world's largest video sharing and search platform.
    \item WeChat\footnote{https://www.wechat.com/} is the most popular social media application in China, supporting sending voice messages and featuring rich Chinese audio materials.
    \item Bilibili\footnote{https://www.bilibili.com/} is a video-sharing website. It has emerged as a major streaming platform in China, offering a wide range of video services.
    \item TikTok\footnote{https://www.tiktok.com/en/} is a short-form video sharing platform. Since its launch, TikTok has become one of the world's most popular social media platforms.
\end{itemize}

$2000$ perturbed audio clips were sourced from the above platforms and applications for manual checking, and $20$ annotators were engaged to annotate the audio clips independently.
All annotators have a Bachelor's degree or higher and are proficient in both English and Chinese.
We provided annotators with substantial guidelines and training regarding the audio classification.
For each audio clip, annotators were required to answer three questions: (1) whether the audio clip is toxic, (2) if the toxic audio is intentionally perturbed to bypass the content moderation software, and (3) how the audio clip is perturbed.
Of the original set, 1,603 clips (80\%) passed verification. These validated clips span all four platforms (YouTube 27.3\%, WeChat 22.9\%, Bilibili 25.7\%, and TikTok 24.1\%), exhibit a near-balanced distribution of English vs. Chinese (54.1\% vs. 45.9\%) and male vs. female speakers (52.4\% vs. 47.6\%), and cover diverse topics (e.g., daily vlogs, pranks, gaming, sports commentary) and scenarios (e.g., live streams, pre-recorded videos, chat dialogues, voice messages).
The label with the highest agreement among annotators was adopted as the final result.
Finally, we manually inspected all the toxic perturbed audio clips and summarized $14$ MRs from the perturbation methods that real users applied to evade the audio moderation as shown in Table \ref{tab:mrs}.

We will introduce these MRs and their corresponding perturbation methods in the following section.

\subsection{MRs}
\label{sec:MRs}

The MRs in MTAM are defined through perturbations applied to the original toxic audios. Concretely, an MR specifies a transformation that modifies the audio while preserving its semantic toxicity label. Based on the implementation form of perturbations, we categorize the MRs into three perspectives: (1) \textbf{Basic Signal-Level Perturbations}: the signal-level audio perturbations manipulate the elementary audio signals such as speed, frequency, amplitude, etc.; (2) \textbf{Compound Signal-Level Perturbations}: perturbations formed by different combinations of basic signal-level perturbations; (3) \textbf{Linguistic-Form Perturbations}: modifications that alter pronunciation or fluency without changing the underlying semantic content. Under basic signal-level perturbations, we summarize five mutually exclusive basic MRs (Section \ref{sec:basis}) related to modifying basic audio signals according to pilot studies.
Then, we identify seven other compound metamorphic methods that modify audio signals (Section \ref{sec:compound}).
Finally, we introduce linguistic-form perturbations with two subcategories found (Section \ref{sec:heuristic}).
Importantly, each MR asserts that after applying the given perturbation, the toxicity label must remain the same.

\subsubsection{\textbf{Basic Signal-Level MRs}}
\label{sec:basis}
These five MRs define basic audio signal perturbations and serve as the fundamental elements for constructing compound signal-level MRs.

\noindent \textbf{MR1-1 Time-Domain Perturbation}

This MR involves directly manipulating the waveform of audio signals over time, which includes changing the tempo of an audio segment by stretching, shifting the segment over time, or periodically perturbing the audio~\cite{Moreira2020TestingAS,Tan2022AdversarialAA,Parascandolo2016RecurrentNN,Salamon2016DeepCN,Wei2020ACO}.

\noindent \textbf{MR1-2 Space-Domain Perturbation}

This MR refers to simulating or manipulating the spatial characteristics of audio~\cite{Mauch2013TheAD,Salamon2016DeepCN}.
For example, stereo panning changes the position in space where the human ear perceives the sound, and surround sound simulates a 3-dimensional rotating sound effect.

\noindent \textbf{MR1-3 Frequency-Domain Perturbation}

This MR employs audio signals transforming to adjust or modify their frequency components~\cite{Wang2022SRMTAM, Tan2022AdversarialAA,Salamon2016DeepCN,Wei2020ACO}.
Pitch shift is the most common manifestation of this MR.

\noindent \textbf{MR1-4 Injection}

This MR perturbs audio via injecting audio signals to the original audio~\cite{peterson2015metamorphic,Moreira2020TestingAS,
Salamon2016DeepCN,Wei2020ACO}.
There are usually two ways of performance, noise injection and repetition.
Noise injection inserts additional waveform into the original sound wave, such as white noise which does not affect human's comprehension.
Repetition refers to create repetition in audio signals to enhance or emphasize certain elements.

\noindent \textbf{MR1-5 Amplitude Adjustment}

This MR refers to modifying the magnitude of signals in an audio segment~\cite{Moreira2020TestingAS}.
This can involve increasing or decreasing the signal's strength to achieve desired loudness, balance, or dynamic range~\cite{Moreira2020TestingAS}.

\subsubsection{\textbf{Compound Signal-Level MRs}}
\label{sec:compound}

According to different combination rules for basic MRs, compound MRs can be divided into the following seven categories:

\noindent \textbf{MR2-1 Compression}

Compression adjusts the amplitude and frequency of the audio signal over time to reduce the difference between the loudest and quietest parts and achieve a balanced output, involving MR1-1, MR1-3, and MR1-5.

\noindent \textbf{MR2-2 Ring Modulation}

Ring modulation manipulates the amplitude of the original signal by multiplying it with another waveform, often a sine or square wave.
As a result, the multiplication of the two signals in ring modulation produces sum and difference frequencies, known as sidebands, that may not exist in the original signal, creating metallic or bell-like tones.
Moreover, the modulation occurs periodically over time, so there’s a time-domain aspect in the interaction between the two waveforms as they are multiplied, which determines the evolution of the signal over time.
Therefore, Ring Modulation also combines of MR1-1, MR1-3, and MR1-5.

\noindent \textbf{MR2-3 Bass Boost}

Bass boost selectively and equally increases the amplitude of low-frequency components, typically 20-200 Hz, to enhance their presence in the mix, consisting of MR1-3 and MR1-5.
MR1-1 could be an optional choice for it to shape or smooth out bass transients.

\noindent \textbf{MR2-4 Tremolo}

This MR adjusts the amplitude and frequency of the audio signal over time, creating a periodic change in volume and pitch to imitate the trembling feeling when people speak with fear.
It is a combination of MR1-1, MR1-3, and MR1-5.

\noindent \textbf{MR2-5 Distortion}

First, this MR clips or compresses the amplitude of the signal to create overtones and alter the perceived loudness and texture of the sound.
Then, it changes the frequency by introducing new harmonics to balance existing frequencies, compromising to the non-linear processing that generates additional frequency components.
During the process, the audio signal evolves over time.
Therefore, distortion is also a mixture of MR1-1, MR1-3, and MR1-5.

\noindent \textbf{MR2-6 Echo}
This MR repeatedly injects delayed original signals with different delays to mimic the effect of echo in the real world, which is a combination of MR1-1 and MR1-4.

\noindent \textbf{MR2-7 Reverb}
This MR injects multiple short delays to create a dense sound field, simulating how sound reflects in a space.
It consists of MR1-2 and MR1-4.

\subsubsection{\textbf{Linguistic-Form MRs}}
\label{sec:heuristic}

MRs under this category are considered "Heuristic", referring to meaning representations derived from surface-level linguistic changes, such as homophone substitution or disfluency that preserve semantic content~\cite{peterson2015metamorphic}.
We have summarized two MRs for this category:

\noindent \textbf{MR3-1 Homophone Substitution}

This MR is based on homophone substitution, which replaces words with other words or character combinations that have the same or similar pronunciation, or slightly changes the accent of an audio segment with similar pronunciation.

Another example is accents in pronunciation.
For instance, pronounce "to" (['tu]) to be "do" (['du]), "color" (['k\textturnv l\textschwa(\textturnr)]) to be "gala" (['ga:l\textschwa]), etc.

In English, simple examples include "folk" ([f\textschwa uk]) for "fuck" ([f\textturnv k]), "sheet" ([\textesh \i \textlengthmark t]) for "shit" ([\textesh \i t]) and "deck" ([d \textepsilon k]) for "dick" ([d\i k]).

In Chinese, the pronunciation of ``\begin{CJK}{UTF8}{gkai}酱\end{CJK}'' ([t\textctc j\textscripta \textipa{N}], Sauce) is similar to that of ``\begin{CJK}{UTF8}{gkai}这样\end{CJK}'' ([t\textrtails \textgamma] [j\textscripta \textipa{N}], Such) when speaking fast~\cite{Wang2023MTTMMT}.
Moreover, Mandarin has four tones, while Cantonese has six tones.
In different regions of China, the same character can be pronounced with different tones.
We can simulate different dialect pronunciations by altering the tones of certain characters.
For example, ``\begin{CJK}{UTF8}{gkai}河\end{CJK}'' (heterophones: [x\textramshorns], River) in Mandarin is the second tone.
However, it pronounces the fourth tone in Henan dialect, a dialect in north China, which is similar to ``\begin{CJK}{UTF8}{gkai}贺\end{CJK}'' (heterophones: [x\textramshorns], Celebrate) in Mandarin.

Additionally, the substitution can happen between different languages. For example, ``exciting'' ([\i k\textprimstress sa\i t\i \textipa{N}]) and ``\begin{CJK}{UTF8}{gkai}亦可赛艇\end{CJK}'' ([\i] [k$^h$\textgamma][sa\i][t\i \textipa{N}], Also/Can/Race/Boat) are acoustically similar~\cite{Wang2023MTTMMT}.

Note that the perturbation logic behind this MR is acoustic semantic similarity rather than textual semantic equivalence.

\noindent \textbf{MR3-2 Benign Discontinuity}

This MR inserts plenty of benign but unrelated acoustic features, such as stops and repetitions, to reduce the toxicity by adding discontinuity.
To be more specific, offensive phrases can be broken down into isolated words, with repeating non-offensive words and long stops between offensive and non-offensive words to imitate stuttering to bypass the hate-speech detection model.

For example, "son of a bitch" could be transformed to "son... son... son...... of a... a... a... bitch" or "fuck" (/f\textturnv k/) could be camouflaged as "/f/.../f//f\textturnv/.../k/.../k/" to bypass moderation in English.

In Chinese, an example is ``\begin{CJK}{UTF8}{gkai}你妈\end{CJK}'' (heterophones: /ni//m\textturnv/, Your mother) could be perturbed as "\begin{CJK}{UTF8}{gkai}你(/ni/, You)...你(/ni/, You)你(/ni/, You)......你(/ni/, You)...妈(/m\textturnv/, Mom)\end{CJK}".

\subsection{Implementation Details}
In this section, we describe the implementation details of \methodname.
More precisely, we implement the perturbation methods in basic signal-level MRs, compound signal-level MRs, and linguistic-form MRs.
To ensure robustness and consistency, all perturbation hyperparameters are carefully tuned via grid search. Each perturbation is required to satisfy the criteria mentioned in Section \ref{sec:pilotstudy}, and has been further validated through human evaluation in Section \ref{RQ1} to confirm that transformed audios remain toxic and comprehensible.

\noindent \textbf{MR1-1 Time-Domain Perturbation.}
The fundamental implementation for this MR is to change the tempo for audio by stretching or compressing it by leveraging a Python library called audiostretchy\footnote{https://pypi.org/project/audiostretchy/}.
Both local and global perturbations are considered for this MR, with more extreme parameters applied in randomized local modification.
When perceived as a component for compound MRs, this MR could extend to include operations that periodically changes or shift the audio by calling corresponding functions in Python over time.

\noindent \textbf{MR1-2 Space-Domain Perturbation.}
There are two applications for this MR: stereo panning and surround.
Stereo panning utilizes Pydub\footnote{https://github.com/jiaaro/pydub} library in Python to call the function "pan" to change audio channels.
Surround overlays different stereo panning by calling the function "overlay" in the same library to simulate a 3-dimensional sound coming from different directions in space.
When perceived as a component for compound MRs, this MR could be abstracted to include operations that modifies the audio to mimic an effect in space.

\noindent \textbf{MR1-3 Frequency-Domain Perturbation.}
MR1-3 uses Pydub, Numpy\footnote{https://github.com/numpy/numpy}, and Scipy\footnote{https://github.com/scipy/scipy} libraries in Python to resample the audio thus shift its frequency and change its pitch.
If perceived as a component for compound MRs, this MR could be generalized to any operations that shifts the frequency of an audio.

\noindent \textbf{MR1-4 Injection.}
We implement two injection methods: noise insertion and repetition.
For noise insertion, we employ Numpy to randomly sample a noise matrix, Librosa\footnote{https://github.com/librosa/librosa} and Soundfile\footnote{https://pypi.org/project/soundfile/} to incorporate the noise matrix into original audio in audio format.
For repetition, we repeat and insert the target audio segments by segmenting and overlaying.

\noindent \textbf{MR1-5 Amplitude Adjustment.}
In this MR, we transform the audio into numerical representation through Pydub library in Python to adjust the amplitude of audio signals by adding or subtracting numbers.
When generalized to a component for compound MRs, this MR could be abstracted to any operations that changes the audio amplitude.

\noindent \textbf{MR2-1 Compression.}
In a Python library Pydub, a function named "compress\_dynamic\_range" can be used to narrow the gap between the maximal volume of a audio segment and the given minimal volume threshold by a given ratio. We use this function to implement the compression MR. 

\noindent \textbf{MR2-2 Ring Modulation.}
To implement this MR, we transform the audio segments to arrays, and multiply them with a sine wave of equal length and given frequency.

\noindent \textbf{MR2-3 Bass Boost.}
We apply the function "low\_pass\_filter" in Pydub to extract the part of audio whose frequency is below a certain cut-off value, and then add this part to a given value in order to boost the bass portion. The bass-boost audio can be ultimately generated by overlaying the same portion of the original audio with the obtained bass portion.

\noindent \textbf{MR2-4 Tremolo.}
Similar to the Ring Modulation MR, we multiply the audio array with a low-frequency sine signal and plus the result with the original audio. Because the frequency of the sine wave is comparatively low, it is not sufficient to cause a frequency change in the audio signal in a short period of time.   

\noindent \textbf{MR2-5 Distortion.}
The processing progress can be divided into three parts: firstly, We clip the portion of the audio that exceeds the amplitude threshold; secondly, we introduce the new harmonics by convolution; finally, a linear variation with respect to time is applied, and we obtain the final distorted audio.

\noindent \textbf{MR2-6 Echo.}
We extract the first short segment of the original audio, and overlay it in a loop to cover the entire original audio. Besides, to make the echo decay, we can introduce an attenuation factor to adjust the volume of the echo.

\noindent \textbf{MR2-7 Reverb.}
Given two parameters, intensity and duration, we can generate a random impulse response array with length of duration and multiply it with intensity. To create the reverb effect, we convolve it with the original audio.

\noindent \textbf{Keyword Selection in MR3:}
In linguistic-form perturbations, we target words critical to content moderation—those frequent in toxic datasets but rare in general corpora—aiming to affect moderation outcomes.
We use TF-IDF to identify such words, applying the sklearn\footnote{https://scikit-learn.org/} library for English and Zuanbot\footnote{https://github.com/cndiandian/zuanbot.com} for Chinese. After removing stop words, we select the top 20 words with the highest TF-IDF scores per dataset.

\noindent \textbf{MR3-1 Homophone Substitution.}
For English, we replace toxic words with phonetically similar alternatives during text-to-speech (TTS) generation, using a curated substitution list; accent variation is also used to induce similar effects. For Chinese, native annotators construct substitution candidates, and the highest-voted alternative replaces the toxic word during TTS. Ties are resolved by random selection.

\noindent \textbf{MR3-2 Benign Discontinuity.}
For English, large-scale implementation is challenging.
For Chinese, we insert full stops and repeated characters to induce non-rhythmic pauses and repetitions in toxic phrases during TTS.
\section{Evaluation}
\label{sec-experiment}

\begin{table}
\centering
\caption{Statistics of Toxic Datasets.}
\begin{tabular}{l r r r r}
\toprule
\bf Dataset & \bf \#Sent & \bf Lang &  \bf Type & \bf Source\\
\midrule
Toxicity & 1.0K  &  English & Abuse   &  Various Platforms \\
Sexting  & 0.5K & English & Porno & Github \\
Zuanbot  &  1.6K & Chinese & Abuse & Github \\
SpamMessage & 60K & Chinese & Spam & Taobao\\
CAnovel  & 1.8M & Chinese & Porno & CAnovel \\
\bottomrule
\end{tabular}
\label{tab:data-statistics}
\end{table}


To evaluate the effectiveness of \methodname, we test our method on five commercial audio moderation software and one audio moderation model.
In this section, we try to answer the following three questions:
\begin{itemize}[leftmargin=*]
    \item Q1: Are the test audios generated by \methodname toxic and realistic?
    \item Q2: Can \methodname find erroneous outputs returned by audio moderation software?
    \item Q3: Can we leverage the test cases created by \methodname to augment the performance of content moderation systems?
\end{itemize}

\subsection{Experimental Settings}

\subsubsection{Datasets}
Different types of datasets are used as seed data in the TTS process to validate \methodname.
In this paper, we select the following datasets:
For English, the Toxicity Dataset\footnote{https://github.com/surge-ai/toxicity} contains 500 toxic and 500 non-toxic comments from a variety of popular social media platforms. The Sexting-dataset\footnote{https://github.com/mathigatti/sexting-dataset} includes pornographic chats scraped from public resources. The dataset of discriminatory, violent, and abusive speech we use comes from this paper \cite{hateoffensive}, containing $24802$ tweets from $33458$ Twitter users. For Chinese, we utilizes CAnovel\footnote{https://canovel.com/}, which is a Chinese pornographic novel website, is a Chinese data source with $3931$ pornographic novels, around 1.8 million pornographic sentences in total, to collect Chinese porn sentences.
The dataset Zuanbot\footnote{https://github.com/cndiandian/zuanbot.com}, which contains $1597$ insult and abusive sentences, is adopted to select desired Chinese abusive and insult sentences.
SpamMessage dataset\footnote{https://github.com/hrwhisper/SpamMessage} is a dataset comprising of around $60K$ real-world Chinese spam messages.
Important statistics of the six datasets are shown in Table~\ref{tab:data-statistics}.


\subsubsection{Software and Models Under Test}
\label{subsec:models}

We use \methodname to test commercial textual content moderation software products and SOTA academic models.
Commercial software products include Assembly AI\footnote{https://www.assemblyai.com/products}, Gladia\footnote{https://www.gladia.io/audio-intelligence}, Baidu AI Cloud\footnote{https://ai.baidu.com/tech/speech/speechcensoring}, Tencent Cloud\footnote{https://www.tencentcloud.com/products/ams}, and Nextdata\footnote{https://intl.ishumei.com/product/audio}, among which Assembly AI and Gladia are for English while Baidu AI Cloud, Tencent Cloud, and Nextdata are for Chinese.
These software products were tested against the three typical kinds of toxic content in our evaluation.
One exception is Tencent Cloud's moderation of insult speeches, since Tencent does not provide such functionality.
The software products listed above are well-known content moderation tools created by various companies, and available through APIs.
For research models, we select models from GitHub and Huggingface\footnote{https://huggingface.co/models} with the highest downloads and stars in the last three years.

\subsection{RQ1: Are the test cases generated by \methodname toxic and realistic?}
\label{RQ1}

The goal of \methodname is to generate toxic and realistic test cases, which mimic how users bypass moderation, based on real-life audio.
In this section, to evaluate their recognizability, toxicity (semantic-preserving), and conformity to real-world users' practices, we applied \methodname to generate $22,071$ perturbed audio clippings and conducted the following test.
we recruited $3$ annotators with a bachelor's degree or higher and proficient in both English and Chinese annotate $2000$ randomly sampled audio pairs without repetition after training, each consisting of a generated test case and its corresponding original audio. For each audio pair, they were asked to score two questions: 
(1) From '$1$ - strongly disagree' to '$5$ - strongly agree', how much do you agree that the perturbed audio is toxic?
(2) From '$1$ - strongly disagree' to '$5$ - strongly agree', how much do you agree that the perturbation applied is consistent with the real-world users' practices in bypassing moderation?
To ensure fairness, annotators reviewed only one shuffled audio at a time, without access to its counterpart, and disagreements were flagged for review.
The annotation results show a mean toxicity score of $4.91$, a recognizability score of $4.29$, and a conformity score to real user behavior of $4.07$.
We follow \cite{Kirk2021HatemojiAT} to measure the inter-rater agreement using Randolph's Kappa, and the obtained value is $0.83$, which indicates 'almost perfect agreement'~\cite{Landis1977TheMO}.

\begin{tcolorbox}[width=\linewidth, boxrule=0pt, colback=gray!20, colframe=gray!20]
\textbf{Answer to RQ1:}
The test cases generated by \methodname are toxic and realistic.
\end{tcolorbox}

\subsection{RQ2: Can \methodname find erroneous outputs returned by audio moderation software?}
\label{subsec-testing-software-with-tool}

\begin{table}
\centering
\caption{Test Cases Statistic.}
\begin{tabular}{l c | r r}
\toprule
\bf Software & \bf Tasks & \bf Ori Num  &\bf Seed Num  \\
\midrule
\multirow{2}{*}{Gladia}& Abuse & 100  &  79\\
 & Porn   & 100  &  60 \\
\midrule
\multirow{2}{*}{Assembly AI}& Abuse & 100  &  88\\
 & Porn   & 100  &  99 \\
\midrule
\multirow{3}{*}{Baidu} & Insult & 100 & 86 \\
 & Porn &  100  &  97\\
& Spam & 100 & 50\\
\midrule
\multirow{3}{*}{NextData}& Insult  & 100 & 96\\
& Porn & 100 & 97\\
&Spam  & 100 & 48 \\
\midrule
\multirow{2}{*}{Tencent}& Porn & 100 & 53\\
&Spam  & 100 & 33 \\
\midrule
\multirow{2}{*}{Academic Model} &  Abuse &100 & 94 \\
 &  Porn & 100 & 71 \\
\bottomrule
\end{tabular}
\label{tab:test_case_stat}
\end{table}

\begin{table*}
\centering
\caption{Error Finding Rates of commercial content moderation software and Academic Models (AM).}
\resizebox{1.0\linewidth}{!}{
\begin{tabular}{l l | l l l l l | l l l l l l | l l l}
\toprule
\multirow{2}{*}{\bf Type} & \multirow{2}{*}{\bf Perturb Methods } & \multicolumn{5}{c}{\bf Insult Detection}   & \multicolumn{6}{c}{\bf Porn Detection}  & \multicolumn{3}{c}{\bf Spam Detection}\\
\cmidrule(lr){3-7} \cmidrule(lr){8-13}  \cmidrule(lr){14-16}  
& &  \bf Gladia & \bf As. AI & \bf Baidu  &\bf NextD. & \bf AM & \bf Gladia & \bf As. AI & \bf Baidu  &\bf NextD. & \bf Tencent & \bf AM   & \bf Baidu  &\bf NextD. & \bf Tencent \\
\midrule
\multirow{4}{*}{Basic}& Speed Change &  3.8  & 1.1 & 36.0 & 24.6 & 54.3 & 1.7 & 2.0  & 39.7 & 10.0 & 40.9 & 97.2 & 49.0 & 3.3 & 50.0 \\
& Pan &0.0 & 0.0 & 50.6 & 0.0 & 0.0 & 0.0 & 0.0 & 19.1 & 0.0 & 0.0 & 0.0 & 31.0 & 0.0 & 23.1 \\
& Pitch Shift & 4.4 & 1.1 & 26.7 & 3.1 & 40.4 & 3.3 & 2.5 & 10.8 & 3.3 & 43.9 & 75.4 & 17.0 & 1.7 & 61.5 \\
& Noise Injection &7.6 & 0.0 & 1.2 & 5.2 & 18.1 & 1.7 & 1.0 & 26.8 & 6.7 & 21.2 & 26.8 & 38.0 & 6.7 & 23.1 \\
\midrule
\multirow{7}{*}{Compound}&Compression & 1.3 &0.0 &9.3 &3.3& 1.1 &0.0 &0.0 &3.1 &0.0& 0.0 &1.4 &4.0& 0.0 &7.7 \\
& Ring Modulation   &  84.8 &75.0 &44.2 &96.9 &58.5 &96.7 &75.8 &63.9 &60.0 &97.0 &98.6 &74.0 &96.7 &100.0   \\
& Bass Boost &79.7 &38.6 &43.0 &26.0 &96.8 &96.7 &35.4 &28.9 &20.0 &63.6 &100.0 &20.0 &3.3 &46.2 \\
& Tremolo &70.9 &8.0 &20.9 &33.3 &29.8 &53.3 &16.2 &18.6 &16.7 &87.9 &66.2 &18.0 &13.3 &61.5 \\
& Distortion & 21.5 &6.8 &100.0 &3.3 &26.6 &15.0 &5.1 &97.9 &3.3 &15.2 &46.5 & 100.0 &0.0 &23.1 \\
& Echo & 21.5 &3.4 &24.4 &7.3 &7.4 &25.0 &22.2 &34.0 &36.7 &66.7 &39.4 &68.0 &23.3 &69.2 \\
& Reverb & 73.4 &21.6 &34.9 &21.9 &12.8 &91.7 &15.2 &11.3 &10.0 &36.4 &45.1 &12.0 &3.3 &46.2 \\
\midrule
\multirow{2}{*}{Heuristic} & Homophone Subs. & 46.2 & 40.4  &39.5 &16.7 &28.7 &35.0 &13.15 &38.1 &13.3 &90.9 &29.6 &30.0 &23.3 &38.5 \\ 
& Benign Dis. & \symbol{92} & \symbol{92} & 31.4 & 0.0 & \symbol{92} & \symbol{92} & \symbol{92} &15.5 &0.0 &39.4 &\symbol{92} &32.0 &3.3 & 46.2 \\
\bottomrule
\end{tabular}
\label{tab:abuse}
}
\end{table*}

\methodname tries to automatically generate test cases to find potential errors in current commercial audio moderation software and academic models.
In this section, we tested this capability on the basis of the number of errors found by \methodname in each audio moderation software.
First, all original audios were input into the tested commercial software and academic model to obtain their moderation labels. Audios labeled as non-toxic by all systems were excluded, as our goal is to find toxic audios that can bypass moderation. The remaining audios served as seed inputs, with statistics shown in Table~\ref{tab:test_case_stat}.
Next, seed audios were perturbed using \methodname’s MRs to generate test cases, which were then re-evaluated by the same systems. We checked whether the test cases retained their original toxicity-related labels (e.g., abusive, pornographic). Since the perturbations preserve semantics, test cases should still be labeled as toxic. Any test case mislabeled as non-toxic indicates a moderation failure.
To evaluate how well \methodname worked on generating test cases resulting errors, the capability of \methodname in this aspect was measured by calculating Error Finding Rate (EFR), which is defined as follows:

$$\text{EFR} = \frac{\text{the number of misclassified test cases}}{\text{the number of generated test cases}} * 100\%.$$

The EFR results are shown in Table~\ref{tab:abuse}. 
In general, \methodname achieves high EFRs.
Compared to academic models, commercial software tends to have lower EFRs.
Specifically, \methodname reaches up to $84.8\%$, $75.9\%$, and $82.5\%$ EFR on commercial products, and up to $100\%$ on academic models.
This gap likely stems from commercial systems adopting rule-based defenses against input perturbations. For instance, Baidu’s patented method for detecting sensitive text\footnote{https://patents.google.com/patent/CN102184188A/en}. 

\begin{tcolorbox}[width=\linewidth, boxrule=0pt, colback=gray!20, colframe=gray!20]
\textbf{Answer to RQ2:}
\methodname achieves $38.6\%$, $18.3\%$, $35.1\%$, $16.7\%$, and $51.1\%$ average error finding rates (EFR) when testing commercial moderation software provided by Gladia, Assembly AI, Baidu, Nextdata, and Tencent respectively, and it obtains up to $49.3\%$ EFR when testing the SOTA acdemic model.
\end{tcolorbox}

\subsection{RQ3: Can we leverage the test cases created by \methodname to augment the performance of content moderation systems?}

We have shown that \methodname is capable of creating toxic and authentic test cases that can bypass the content moderation of commercial software and the academic models.
As shown in the "Insult Detection" column of Table~\ref{tab:abuse}, \methodname achieves high EFRs on academic models across most MRs, suggesting these models are easily fooled.
This raises an important question: can these test cases be used to enhance the performance of content moderation? To put it another way, our goal is to improve the models' robustness.
A potential solution is to retrain the models using the test cases generated by \methodname and assess whether the retrained models exhibit greater resilience to various perturbations.

Concretely, the dataset includes $100$ porn, $100$ insult, and $100$ non-toxic audios for each of the 12 basic and compound MRs (except Pitch Shift and Homophone Substitution). For Pitch Shift, we generate ascending and descending versions. For Homophone Substitution, we apply three variants: phonetic substitutions, and Indian/Singaporean-accented speech, each with $100$ samples per class.

We split the dataset into training and test sets. The test set includes $20\%$ of the data with balanced classes per MR. The training set also includes $20\%$ of the total, with equal class distribution.

We then fine-tune the pre-trained Roblox Voice Safety Model~\cite{roblox_voice_safety_classifier} using default settings\footnote{\url{https://huggingface.co/Roblox/voice-safety-classifier}}, without modifying any hyperparameters.


\begin{table}
\centering
\caption{Error Finding Rates (EFRs) on models after retraining on $2$ training set and the original model.}
\resizebox{1.0\linewidth}{!}{%
\begin{tabular}{l l | c c c | c c c}
\toprule
\bf \multirow{2}{*}{Type} & \bf \multirow{2}{*}{Perturb Methods} & \multicolumn{3}{c|}{\bf Ori} & \multicolumn{3}{c}{\bf Aug}\\
\cmidrule(lr){3-5} \cmidrule(lr){6-8}
& & \bf Insult & \bf Porn & \bf Non-toxic & \bf Insult & \bf Porn & \bf Non-toxic\\
\midrule
\multirow{4}{*}{Basic}
& Speed Change    & 25.0  & 50.0  & 0.0  & 0.0 & 5.0  & 0.0  \\
& Pan             & 20.0  & 35.0  & 0.0  & 0.0 & 10.0  & 0.0  \\
& Pitch Shift     & 60.0  & 77.5  & 0.0  & 7.5 & 2.5  & 7.5  \\
& Noise Injection & 35.0  & 60.0  & 0.0  & 0.0 & 10.0  & 0.0  \\
\midrule
\multirow{7}{*}{Compound}
& Compression     & 20.0  & 35.0  & 0.0  & 0.0 & 10.0  & 0.0  \\
& Ring Modulate   & 95.0  & 100.0  & 5.0  & 5.0 & 15.0  & 10   \\
& Bass Boost      & 100.0 & 100.0  & 0.0  & 40 & 5.0  & 0.0   \\
& Tremolo         & 40.0  & 80.0  & 5.0  & 5.0 & 10.0  & 5.0  \\
& Distortion      & 60.0  & 65.0  & 0.0  & 5.0 & 0.0  & 5.0   \\
& Echo            & 20.0  & 60.0  & 0.0  & 0.0 & 5.0  & 5.0   \\
& Reverb          & 35.0  & 50.0  & 5.0  & 5.0 & 10.0  & 0.0   \\
\midrule
\multirow{1}{*}{Heuristic}
& Homophone Subs. & 95.0 & 41.7 & 1.7 & 11.7 & 1.7 & 6.7  \\
\bottomrule
\end{tabular}
}
\label{tab:abuse_improve}
\end{table}

The results in Table~\ref{tab:abuse_improve} indicate that the test cases generated by \methodname significantly enhance the robustness of audio moderation models, as evidenced by the considerable reduction in EFRs.
Meanwhile, after retraining with \methodname's test cases, the model is never deceived by all perturbations.
Furthermore, we also notice that the performance of the model retraining on the training set of $20\%$ data is comparably robust enough, which means only a small amount of data is needed to remarkably enhance the robustness of the model.
We also note a slight decrease in the accuracy of the retrained models for non-toxic audio classification, but fortunately this dip in performance appears acceptable compared to the rise in accuracy for toxic audio classification.

Regrettably, experiments to enhance industrial models are not conducted, as industrial moderation software typically operates as black boxes with only accessible APIs, while retraining necessitates complete access to the model kernel.
Nevertheless, we posit that retraining using \methodname's test cases would similarly bolster the robustness of industrial models, given the comparable underlying models.

\begin{tcolorbox}[width=\linewidth, boxrule=0pt, colback=gray!20, colframe=gray!20]
\textbf{Answer to RQ3:}
Fine-tuning with MTAM test cases led to improvements in the Error Finding Rates (EFR) across all evaluated models.
\end{tcolorbox}

\subsection{Compared with Audio Adversarial Attack Methods}

\begin{table}
\centering
\caption{Comparison of Adversarial Attack Methods and \methodname on EFR in English (EN) and Chinese (ZH), Query Times, and Applicability for Academic Models.}
\resizebox{0.48\textwidth}{!}{%
\begin{tabular}{l c | c c | c | c}
\toprule
\bf Attack Type & \bf Methods & \bf{EN EFR} & \bf{ZH EFR} & \bf Queries & \bf Applicability\\
\midrule
\multirow{2}{*}{White-Box}& FGSM & 79.1 &  85.6 & 4.0 & Limited \\
 & PGD   & \bf 90.5 & \bf 88.5 & 10.0 & Limited  \\
\midrule
\multirow{3}{*}{Black-Box}& Boundary & 35.0 & 29.9 & 10.0 & Vast \\
 & ZOO   & 10.5 &  25.3 & 24.8 & Vast \\
 \cmidrule(lr){2-6}
 &\methodname (ours) & \bf{57.2} & \bf{41.4} & \bf 1.0 & \bf Vast \\
\bottomrule
\end{tabular}%
}
\label{tab:advCompare_stat}
\end{table}

In this section, we will illustrate the advantage of \methodname compared to another research direction for discovering errors in audio moderation software and audio adversarial attack methods.

First, the dominating adversarial attack methods are white-box methods nowadays~\cite{Tan2022AdversarialAA}, with only a few black-box approaches available.
In this case, attackers often exploit the detailed knowledge of the model's architecture, parameters, and gradients to generate adversarial examples efficiently.
However, most commercial audio moderation software is not open source, making our comparison difficult.

Second, \methodname is more comprehensive than single audio adversarial methods because most of these methods mainly concentrate on a small subset of the perturbations in \methodname (e.g. Gaussian noise\cite{khare2018adversarial}) as our metamorphic relations can generalize most black-box adversarial attack strategies~\cite{Tan2022AdversarialAA}.

To demonstrate the effectiveness of \methodname, we conducted an experiment comparing the performance of \methodname with audio adversarial attack methods in terms of EFR and runtime. Concretely, we attacked the corresponding academic models mentioned above in English and Chinese using four mainstream adversarial attack methods: Fast Gradient Sign Method (FGSM)~\cite{Goodfellow2014ExplainingAH}, Projected Gradient Descent (PGD)~\cite{Madry2017TowardsDL}, Boundary Attack~\cite{Brendel2017DecisionBasedAA}, and Zeroth Order Optimization (ZOO) Attack~\cite{Chen2017ZOOZO}.
FGSM and PGD are white-box, and Boundary Attack and ZOO Attack are black-box.

Specifically, In FGSM and PGD, we set $\epsilon \leq 0.1$, which is a small value we multiply the signed gradients by to ensure the perturbations are small enough that do not affect human recognition but large enough that they fool the model, and additionally for PGD, the number of iterations is set as $10$.
In the boundary attack and ZOO attack, we take the maximum number of iterations as $10$, to align with the PGD attack and control the noise magnitude.

In general, according to Table~\ref{tab:advCompare_stat}, white-box adversarial attack methods outperform \methodname regardless of language.
However, they require full access to and complete knowledge of the models.
Additionally, more model query times are needed than \methodname.
For black-box methods, \methodname significantly outperforms Boundary Attack and ZOO Attack in terms of both EFR and runtime, irrespective of language.

Explicitly, for the English model, the decreases in model confidence in identifying a certain class are, at average, 60.0\% and 92.9\% in terms of FGSM and PGD attacks separately, which causes loads of misclassifications and is much better than \methodname.
Moreover, the black-box Boundary Attack demonstrates an average 24.8\% decrease in the classifying confidence, and the maximum and minimum decreases are 86.9\% and -40.9\%, separately.
This method shows limited effectiveness as the EFR is 35\%, generally lower than our method.
The runtime efficiency of \methodname lies in the minimum model query times of only once, while the ZOO attack requires $24.8$ model query times on average, and the Boundary attack demands an entire $10$ times queries.
For the Chinese binary classifier, \methodname outperforms Boundary Attack and ZOO attack, as the average EFR of Boundary Attack is $29.9\%$ when audios sound similar to the audios under MR1-4 Injection and $25.3\%$ for ZOO attack.

Notably, regardless of language type, most of the perturbation methods devised by our MRs, which are above average, are even much more effective than the black-box adversarial attacks. For example, \textit{Bass Boost} has an EFR of $96.8\%$ according to table~\ref{tab:abuse}, outperforms both Boundary Attack and ZOO Attack.

In summary, \methodname can find more errors in less running time.
\section{Related Works}

\subsection{Robustness of AI Software}

\textit{AI software} has seen widespread adoption across various domains, including autonomous driving and facial recognition. Despite this, the robustness of AI systems remains a significant challenge, as they can sometimes produce erroneous results, leading to potentially catastrophic consequences, such as fatal accidents \cite{notrobustself-driving,notrobusttesla}. In response to these issues, numerous methods have been developed to create adversarial examples or test cases aimed at tricking AI systems~\cite{Carlini2016HiddenVC, Tu2021ExploringAR, Luo2021InteractivePF, Pei2017DeepXploreAW, Zhang2022MachineLT}.

Various strategies have emerged to generate adversarial test cases, targeting the vulnerabilities of AI software \cite{Carlini2016HiddenVC, Tu2021ExploringAR, Luo2021InteractivePF, Pei2017DeepXploreAW, Zhang2022MachineLT}. Concurrently, research efforts have focused on enhancing the robustness of AI systems, such as through robust training methods~\cite{Madry2018TowardsDL, Asyrofi2021CanDT, Gao2020FuzzTB} and network debugging techniques~\cite{Ma2018MODEAN,Tao2020TRADERTD}.

In recent years, NLP software has also experienced significant advancements. It is now commonly applied in tasks like sentiment analysis~\cite{Zhang2017SentimentAA,Wang2017EmotionRW}, machine translation~\cite{Bahdanau2015NeuralMT,Wang2022UnderstandingAI,Jiao2022TencentsMM}, and text-to-speech synthesis~\cite{Wang2017TacotronTE, Ma2018FPETSFP}. Given its increasing importance, researchers from both the NLP and software engineering communities have begun examining the robustness of NLP models~\cite{Gupta2020MachineTT, He2021TestingMT, Jiao2023IsCA}. For example, Ribeiro et al.\cite{Ribeiro2020BeyondAB} introduced a behavioral testing framework that evaluates NLP software used for sentiment analysis, question answering, and machine comprehension. Li et al.\cite{Li2020BERTATTACKAA} employed deep learning techniques to generate test cases targeting NLP models, while Sun et al.~\cite{Sun2022ImprovingMT} proposed a word-replacement method to address machine translation errors without retraining the model.

In our study, we focus on analyzing the robustness of content moderation software, a widely-used AI tool that has not been systematically investigated before.

\subsection{Robustness of Audio Moderation Software}

We systematically reviewed papers on metamorphic relations, audio processing, and evaluating as well as challenging content moderation within related fields such as software engineering, natural language processing, and speech signal processing.
Most audio moderation software nowadays adopt the SOTA audio processing methods such as ASR based on neural networks\cite{jain2020contextual}.
Powered by advanced NLP techniques, these software have made enormous breakthroughs in moderation accuracy.
There are solid papers discussing the techniques and MRs mentioned in our work.
Specifically, Moreira et al.~\cite{Moreira2020TestingAS} defined similar metamorphic relations such as amplitude (MR1-5), noise insertion (MR1-4), and shift(MR1-1) to test acoustic scene classifiers.
Tan et al.~\cite{Tan2022AdversarialAA} introduced Time-Domain Perturbation (MR1-1) and Frequency-Domain Perturbation (MR1-3) as basic perturbation objects.
Mauch and Ewert~\cite{Mauch2013TheAD} involved speed-up (MR1-1) as an audio degradation toolbox, serving as robustness evaluation approach.
Parascandolo et al.~\cite{Parascandolo2016RecurrentNN} mentioned time-domain augmentation method (MR1-1) in experiments.
Salamon and Bello~\cite{Salamon2016DeepCN} discussed frequency-domain perturbation (MR1-3) and compression (MR2-1-1) in data augmentation for Environmental Sound Classification.
Wei et al.~\cite{Wei2020ACO} found noise injection (MR1-4) also effective in data augmentation for audio classification.

However, our paper makes a substantial contribution in comparison to the aforementioned studies.
First, \methodname is much more comprehensive.
Our MRs not only summarizes the perturbation methods stated in previous work, but also categorizes the metamorphic relations on audio that are common in daily lives.
To the best of our knowledge, the other MRs in \methodname have not been investigated in the current literature across various related fields.
Moreover, all these papers focus on English setting, while we also consider \methodname for Chinese.
Additionally, all the MRs were backed by our pilot study involving actual user inputs, unlike existing papers that relied on perturbations derived from domain knowledge.
Furthermore, we extended the evaluation on five commercial audio moderation software products rather than most of the existing papers which only evaluate on research models.
Therefore, we believe that \methodname is the first thorough testing framework for audio moderation.

\section{Conclusion}

This paper introduces \methodname, a comprehensive testing framework designed to validate audio moderation software.
In contrast to existing testing methods or adversarial attack techniques for general NLP applications, which mainly offer basic perturbations and address only a limited range of toxic inputs that malicious users might create, \methodname includes eleven metamorphic relations primarily inspired by a preliminary study and previous research.
Furthermore, all these metamorphic relations have been implemented for both English and Chinese.
Our evaluation indicates that the test cases generated by \methodname can effectively bypass the moderation of six state-of-the-art moderation algorithms as well as commercial content moderation tools from Tencent, Baidu, Nextdata, Assembly AI and Gladia.
These test cases have been used to retrain the algorithms, resulting in significant improvements in model robustness while maintaining the same accuracy on the original test set.
We view this work as a vital first step towards systematic testing of audio moderation software.
In future work, we plan to further develop the metamorphic relations in \methodname and extend its application to additional languages.
Additionally, we aim to initiate a comprehensive effort to continuously test and enhance content moderation software.

\section{Acknowledgement}
The work described in this paper was supported by the Research Grants Council of the Hong Kong Special Administrative Region, China (No. CUHK 14206921 of the General Research Fund) and the National Natural Science Foundation of China (Grant Nos. 62102340 and 62206318). The HKUST authors are supported in part by a RGC GRF grant under the contract 16214723.

\balance
\bibliographystyle{IEEEtran}
\bibliography{reference}

\begin{thebibliography}{10}
\providecommand{\url}[1]{#1}
\csname url@samestyle\endcsname
\providecommand{\newblock}{\relax}
\providecommand{\bibinfo}[2]{#2}
\providecommand{\BIBentrySTDinterwordspacing}{\spaceskip=0pt\relax}
\providecommand{\BIBentryALTinterwordstretchfactor}{4}
\providecommand{\BIBentryALTinterwordspacing}{\spaceskip=\fontdimen2\font plus
\BIBentryALTinterwordstretchfactor\fontdimen3\font minus \fontdimen4\font\relax}
\providecommand{\BIBforeignlanguage}[2]{{%
\expandafter\ifx\csname l@#1\endcsname\relax
\typeout{** WARNING: IEEEtran.bst: No hyphenation pattern has been}%
\typeout{** loaded for the language `#1'. Using the pattern for}%
\typeout{** the default language instead.}%
\else
\language=\csname l@#1\endcsname
\fi
#2}}
\providecommand{\BIBdecl}{\relax}
\BIBdecl

\bibitem{tiktok2024}
B.~Team, ``Tiktok statistics you need to know,'' \url{https://backlinko.com/tiktok-users}, 2024, accessed: 2024-09-14.

\bibitem{Ren2020FastSpeech2F}
Y.~Ren, C.~Hu, X.~Tan, T.~Qin, S.~Zhao, Z.~Zhao, and T.-Y. Liu, ``Fastspeech 2: Fast and high-quality end-to-end text to speech,'' \emph{ArXiv}, vol. abs/2006.04558, 2020.

\bibitem{Badjatiya2017DeepLF}
P.~Badjatiya, S.~Gupta, M.~Gupta, and V.~Varma, ``Deep learning for hate speech detection in tweets,'' \emph{Proceedings of the 26th International Conference on World Wide Web Companion}, 2017.

\bibitem{Rowley2006LargeSI}
H.~A. Rowley, Y.~Jing, and S.~Baluja, ``Large scale image-based adult-content filtering,'' in \emph{VISAPP}, 2006.

\bibitem{Li2012KnowingYE}
Z.~Li, K.~Zhang, Y.~Xie, F.~Yu, and X.~Wang, ``Knowing your enemy: understanding and detecting malicious web advertising,'' \emph{Proceedings of the 2012 ACM conference on Computer and communications security}, 2012.

\bibitem{children2011}
E.~R. Munro, ``The protection of children online: a brief scoping review to identify vulnerable groups,'' \emph{Childhood Wellbeing Research Centre}, 2011.

\bibitem{Yu2016InternetMI}
T.-K. Yu and C.-M. Chao, ``Internet misconduct impact adolescent mental health in taiwan: The moderating roles of internet addiction,'' \emph{International Journal of Mental Health and Addiction}, vol.~14, pp. 921--936, 2016.

\bibitem{spam2022}
N.~Cveticanin, ``What's on the other side of your inbox - 20 spam statistics for 2022,'' \url{https://dataprot.net/statistics/spam-statistics/}, 2022, accessed: 2022-03-01.

\bibitem{Chen2020AutomaticDO}
Y.~Chen, R.~Zheng, A.~Zhou, S.~Liao, and L.~Liu, ``Automatic detection of pornographic and gambling websites based on visual and textual content using a decision mechanism,'' \emph{Sensors (Basel, Switzerland)}, vol.~20, 2020.

\bibitem{Mishra2019TacklingOA}
P.~Mishra, H.~Yannakoudakis, and E.~Shutova, ``Tackling online abuse: A survey of automated abuse detection methods,'' \emph{ArXiv}, vol. abs/1908.06024, 2019.

\bibitem{Schmidt2017ASO}
A.~Schmidt and M.~Wiegand, ``A survey on hate speech detection using natural language processing,'' in \emph{SocialNLP@EACL}, 2017.

\bibitem{Wu2018TwitterSD}
T.~Wu, S.~Wen, Y.~Xiang, and W.~Zhou, ``Twitter spam detection: Survey of new approaches and comparative study,'' \emph{Comput. Secur.}, vol.~76, pp. 265--284, 2018.

\bibitem{Gupta2022ADIMAAD}
V.~Gupta, R.~A. Sharon, R.~Sawhney, and D.~Mukherjee, ``Adima: Abuse detection in multilingual audio,'' \emph{ICASSP 2022 - 2022 IEEE International Conference on Acoustics, Speech and Signal Processing (ICASSP)}, pp. 6172--6176, 2022.

\bibitem{Devlin2019BERTPO}
J.~Devlin, M.-W. Chang, K.~Lee, and K.~Toutanova, ``Bert: Pre-training of deep bidirectional transformers for language understanding,'' \emph{NAACL}, vol. abs/1810.04805, 2019.

\bibitem{Liu2019RoBERTaAR}
Y.~Liu, M.~Ott, N.~Goyal, J.~Du, M.~Joshi, D.~Chen, O.~Levy, M.~Lewis, L.~Zettlemoyer, and V.~Stoyanov, ``Roberta: A robustly optimized bert pretraining approach,'' \emph{ArXiv}, vol. abs/1907.11692, 2019.

\bibitem{notrobustbaidu}
M.~Jing, ``China’s baidu turns to ai to police online content, but is the technology reliable?'' \url{https://www.scmp.com/tech/innovation/article/2143759/chinas-baidu-turns-ai-police-online-content-technology-reliable?module=perpetual_scroll_0&pgtype=article&campaign=2143759}, 2018, accessed: 2022-03-01.

\bibitem{tenaudmod2025}
\BIBentryALTinterwordspacing
Tencent, ``Audio moderation system,'' 2025, accessed: 2025-01-10. [Online]. Available: \url{https://www.tencentcloud.com/products/ams}
\BIBentrySTDinterwordspacing

\bibitem{glaAudmod2025}
\BIBentryALTinterwordspacing
Gladia, ``Content moderation,'' 2025, accessed: 2025-01-10. [Online]. Available: \url{https://docs.gladia.io/chapters/audio-intelligence/pages/moderation}
\BIBentrySTDinterwordspacing

\bibitem{bypassMod2022}
S.~Walker, ``9 sneaky ways people bypass auto-moderation,'' \url{https://newmediaservices.com.au/9-ways-to-bypass-auto-moderation/}, 2022, accessed: 2024-09-15.

\bibitem{Radford2022RobustSR}
A.~Radford, J.~W. Kim, T.~Xu, G.~Brockman, C.~McLeavey, and I.~Sutskever, ``Robust speech recognition via large-scale weak supervision,'' \emph{ArXiv}, vol. abs/2212.04356, 2022.

\bibitem{AEON2022ISSTA}
J.~Huang, J.~Zhang, W.~Wang, P.~He, Y.~Su, and M.~R. Lyu, ``{AEON:} a method for automatic evaluation of {NLP} test cases,'' in \emph{International Symposium on Software Testing and Analysis (ISSTA)}, 2022.

\bibitem{Wang2023SoftwareTW}
J.~Wang, Y.~Huang, C.~Chen, Z.~Liu, S.~Wang, and Q.~Wang, ``Software testing with large language models: Survey, landscape, and vision,'' \emph{IEEE Transactions on Software Engineering}, vol.~50, pp. 911--936, 2023.

\bibitem{Zhang2014SearchbasedIO}
J.~M. Zhang, J.~Chen, D.~Hao, Y.~Xiong, B.~Xie, L.~Zhang, and H.~Mei, ``Search-based inference of polynomial metamorphic relations,'' \emph{Proceedings of the 29th ACM/IEEE International Conference on Automated Software Engineering}, 2014.

\bibitem{Zhang2019AutomaticDA}
B.~Zhang, H.~Zhang, J.~Chen, D.~Hao, and P.~Moscato, ``Automatic discovery and cleansing of numerical metamorphic relations,'' \emph{2019 IEEE International Conference on Software Maintenance and Evolution (ICSME)}, pp. 235--245, 2019.

\bibitem{Carzaniga2014CrosscheckingOF}
A.~Carzaniga, A.~Goffi, A.~Gorla, A.~Mattavelli, and M.~Pezz{\`e}, ``Cross-checking oracles from intrinsic software redundancy,'' \emph{Proceedings of the 36th International Conference on Software Engineering}, 2014.

\bibitem{Malik2020AutomaticSR}
\BIBentryALTinterwordspacing
M.~Malik, M.~K. Malik, K.~Mehmood, and I.~Makhdoom, ``Automatic speech recognition: a survey,'' \emph{Multimedia Tools and Applications}, vol.~80, pp. 9411 -- 9457, 2020. [Online]. Available: \url{https://api.semanticscholar.org/CorpusID:228864901}
\BIBentrySTDinterwordspacing

\bibitem{ittichaichareon2012speech}
C.~Ittichaichareon, S.~Suksri, and T.~Yingthawornsuk, ``Speech recognition using mfcc,'' in \emph{International conference on computer graphics, simulation and modeling}, vol.~9, 2012.

\bibitem{swietojanski2013revisiting}
P.~Swietojanski, A.~Ghoshal, and S.~Renals, ``Revisiting hybrid and gmm-hmm system combination techniques,'' in \emph{2013 IEEE International Conference on Acoustics, Speech and Signal Processing}.\hskip 1em plus 0.5em minus 0.4em\relax IEEE, 2013, pp. 6744--6748.

\bibitem{jain2020contextual}
M.~Jain, G.~Keren, J.~Mahadeokar, G.~Zweig, F.~Metze, and Y.~Saraf, ``Contextual rnn-t for open domain asr,'' \emph{arXiv preprint arXiv:2006.03411}, 2020.

\bibitem{GoogleMod2023}
E.~K. Colby~Hawker, ``Improving trust in ai and online communities with palm-based moderation,'' \url{https://cloud.google.com/blog/products/ai-machine-learning/google-cloud-text-moderation/}, 2023, accessed: 2024-09-15.

\bibitem{BaiduMod2024}
Baidu, ``Speech technology,'' \url{https://intl.cloud.baidu.com/product/speech.html}, 2024, accessed: 2024-09-15.

\bibitem{TencentMod2024}
T.~Cloud, ``Content moderation,'' \url{https://www.tencentcloud.com/document/product/436/53946}, 2024, accessed: 2024-09-15.

\bibitem{Djuric2015HateSD}
N.~Djuric, J.~Zhou, R.~Morris, M.~Grbovic, V.~Radosavljevic, and N.~L. Bhamidipati, ``Hate speech detection with comment embeddings,'' \emph{Proceedings of the 24th International Conference on World Wide Web}, 2015.

\bibitem{pennington2014glove}
J.~Pennington, R.~Socher, and C.~D. Manning, ``Glove: Global vectors for word representation,'' in \emph{Proceedings of the 2014 conference on empirical methods in natural language processing (EMNLP)}, 2014, pp. 1532--1543.

\bibitem{Chen2020MetamorphicTA}
T.~Y. Chen, S.~C. Cheung, and S.-M. Yiu, ``Metamorphic testing: A new approach for generating next test cases,'' \emph{ArXiv}, vol. abs/2002.12543, 2020.

\bibitem{10.1145/3425174.3425226}
S.~a. H.~N. Santos, B.~N.~C. da~Silveira, S.~a.~A. Andrade, M.~Delamaro, and S.~R.~S. Souza, ``An experimental study on applying metamorphic testing in machine learning applications,'' in \emph{Proceedings of the 5th Brazilian Symposium on Systematic and Automated Software Testing}, ser. SAST '20.\hskip 1em plus 0.5em minus 0.4em\relax New York, NY, USA: Association for Computing Machinery, 2020, p. 98–106.

\bibitem{9477683}
Y.~Deng, G.~Lou, X.~Zheng, T.~Zhang, M.~Kim, H.~Liu, C.~Wang, and T.~Y. Chen, ``Bmt: Behavior driven development-based metamorphic testing for autonomous driving models,'' in \emph{2021 IEEE/ACM 6th International Workshop on Metamorphic Testing (MET)}, 2021, pp. 32--36.

\bibitem{JIANG2022106966}
M.~Jiang, T.~Y. Chen, and S.~Wang, ``On the effectiveness of testing sentiment analysis systems with metamorphic testing,'' \emph{Information and Software Technology}, vol. 150, p. 106966, 2022.

\bibitem{10.1145/3551349.3561157}
\BIBentryALTinterwordspacing
Y.~Yuan, Q.~Pang, and S.~Wang, ``Unveiling hidden dnn defects with decision-based metamorphic testing,'' in \emph{Proceedings of the 37th IEEE/ACM International Conference on Automated Software Engineering}, ser. ASE '22.\hskip 1em plus 0.5em minus 0.4em\relax New York, NY, USA: Association for Computing Machinery, 2023. [Online]. Available: \url{https://doi.org/10.1145/3551349.3561157}
\BIBentrySTDinterwordspacing

\bibitem{Dwarakanath2018IdentifyingIB}
A.~Dwarakanath, M.~Ahuja, S.~Sikand, R.~M. Rao, R.~P. J.~C. Bose, N.~Dubash, and S.~Podder, ``Identifying implementation bugs in machine learning based image classifiers using metamorphic testing,'' \emph{Proceedings of the 27th ACM SIGSOFT International Symposium on Software Testing and Analysis}, 2018.

\bibitem{peterson2015metamorphic}
M.~Peterson, ``Metamorphic testing of sensor processing for android applications,'' Ph.D. dissertation, Virginia State University, 2015.

\bibitem{Moreira2020TestingAS}
D.~D. Moreira, A.~P. Furtado, and S.~C. Nogueira, ``Testing acoustic scene classifiers using metamorphic relations,'' \emph{2020 IEEE International Conference On Artificial Intelligence Testing (AITest)}, pp. 47--54, 2020.

\bibitem{Wang2022SRMTAM}
F.~Wang, K.~Ben, and X.~Zhang, ``Sr-mt: A metamorphic method to test the robustness of speech recognition software,'' \emph{2022 IEEE/ACM 7th International Workshop on Metamorphic Testing (MET)}, pp. 15--22, 2022.

\bibitem{Tan2022AdversarialAA}
\BIBentryALTinterwordspacing
H.~Tan, L.~Wang, H.~Zhang, J.~Zhang, M.~Shafiq, and Z.~Gu, ``Adversarial attack and defense strategies of speaker recognition systems: A survey,'' \emph{Electronics}, 2022. [Online]. Available: \url{https://api.semanticscholar.org/CorpusID:250541764}
\BIBentrySTDinterwordspacing

\bibitem{Mauch2013TheAD}
\BIBentryALTinterwordspacing
M.~Mauch and S.~Ewert, ``The audio degradation toolbox and its application to robustness evaluation,'' in \emph{International Society for Music Information Retrieval Conference}, 2013. [Online]. Available: \url{https://api.semanticscholar.org/CorpusID:11675708}
\BIBentrySTDinterwordspacing

\bibitem{Parascandolo2016RecurrentNN}
\BIBentryALTinterwordspacing
G.~Parascandolo, H.~Huttunen, and T.~Virtanen, ``Recurrent neural networks for polyphonic sound event detection in real life recordings,'' \emph{2016 IEEE International Conference on Acoustics, Speech and Signal Processing (ICASSP)}, pp. 6440--6444, 2016. [Online]. Available: \url{https://api.semanticscholar.org/CorpusID:1810645}
\BIBentrySTDinterwordspacing

\bibitem{Salamon2016DeepCN}
\BIBentryALTinterwordspacing
J.~Salamon and J.~P. Bello, ``Deep convolutional neural networks and data augmentation for environmental sound classification,'' \emph{IEEE Signal Processing Letters}, vol.~24, pp. 279--283, 2016. [Online]. Available: \url{https://api.semanticscholar.org/CorpusID:3537408}
\BIBentrySTDinterwordspacing

\bibitem{Wei2020ACO}
\BIBentryALTinterwordspacing
S.~Wei, S.~Zou, F.~Liao, and W.~Lang, ``A comparison on data augmentation methods based on deep learning for audio classification,'' \emph{Journal of Physics: Conference Series}, vol. 1453, 2020. [Online]. Available: \url{https://api.semanticscholar.org/CorpusID:215994154}
\BIBentrySTDinterwordspacing

\bibitem{Wang2023MTTMMT}
W.~Wang, J.~tse Huang, W.~Wu, J.~Zhang, Y.~Huang, S.~Li, P.~He, and M.~R. Lyu, ``Mttm: Metamorphic testing for textual content moderation software,'' \emph{2023 IEEE/ACM 45th International Conference on Software Engineering (ICSE)}, pp. 2387--2399, 2023.

\bibitem{hateoffensive}
T.~Davidson, D.~Warmsley, M.~Macy, and I.~Weber, ``Automated hate speech detection and the problem of offensive language,'' in \emph{Proceedings of the 11th International AAAI Conference on Web and Social Media}, ser. ICWSM '17, 2017, pp. 512--515.

\bibitem{Kirk2021HatemojiAT}
H.~R. Kirk, B.~Vidgen, P.~R{\"o}ttger, T.~Thrush, and S.~A. Hale, ``Hatemoji: A test suite and adversarially-generated dataset for benchmarking and detecting emoji-based hate,'' \emph{ACL}, vol. abs/2108.05921, 2021.

\bibitem{Landis1977TheMO}
\BIBentryALTinterwordspacing
J.~R. Landis and G.~G. Koch, ``The measurement of observer agreement for categorical data.'' \emph{Biometrics}, vol. 33 1, pp. 159--74, 1977. [Online]. Available: \url{https://api.semanticscholar.org/CorpusID:11077516}
\BIBentrySTDinterwordspacing

\bibitem{roblox_voice_safety_classifier}
\BIBentryALTinterwordspacing
Roblox, ``Voice safety classifier,'' 2024, accessed: 2024-12-13. [Online]. Available: \url{https://huggingface.co/Roblox/voice-safety-classifier}
\BIBentrySTDinterwordspacing

\bibitem{khare2018adversarial}
S.~Khare, R.~Aralikatte, and S.~Mani, ``Adversarial black-box attacks on automatic speech recognition systems using multi-objective evolutionary optimization,'' \emph{arXiv preprint arXiv:1811.01312}, 2018.

\bibitem{Goodfellow2014ExplainingAH}
\BIBentryALTinterwordspacing
I.~J. Goodfellow, J.~Shlens, and C.~Szegedy, ``Explaining and harnessing adversarial examples,'' \emph{CoRR}, vol. abs/1412.6572, 2014. [Online]. Available: \url{https://api.semanticscholar.org/CorpusID:6706414}
\BIBentrySTDinterwordspacing

\bibitem{Madry2017TowardsDL}
A.~Madry, A.~Makelov, L.~Schmidt, D.~Tsipras, and A.~Vladu, ``Towards deep learning models resistant to adversarial attacks,'' \emph{ArXiv}, vol. abs/1706.06083, 2017.

\bibitem{Brendel2017DecisionBasedAA}
W.~Brendel, J.~Rauber, and M.~Bethge, ``Decision-based adversarial attacks: Reliable attacks against black-box machine learning models,'' \emph{ArXiv}, vol. abs/1712.04248, 2017.

\bibitem{Chen2017ZOOZO}
\BIBentryALTinterwordspacing
P.-Y. Chen, H.~Zhang, Y.~Sharma, J.~Yi, and C.-J. Hsieh, ``Zoo: Zeroth order optimization based black-box attacks to deep neural networks without training substitute models,'' \emph{Proceedings of the 10th ACM Workshop on Artificial Intelligence and Security}, 2017. [Online]. Available: \url{https://api.semanticscholar.org/CorpusID:2179389}
\BIBentrySTDinterwordspacing

\bibitem{notrobustself-driving}
C.~Ziegler, ``A google self-driving car caused a crash for the first time. [online],'' \url{https://www.theverge.com/2016/2/29/11134344/google-self-driving-car-crash-report}, 2016, accessed: 2016-09.

\bibitem{notrobusttesla}
S.~Levin, ``Tesla fatal crash: 'autopilot' mode sped up car before driver killed, report finds [online],'' \url{https://www.theguardian.com/technology/2018/jun/07/tesla-fatal-crash-silicon-valley-autopilot-mode-report}, 2018, accessed: 2018-06.

\bibitem{Carlini2016HiddenVC}
N.~Carlini, P.~Mishra, T.~Vaidya, Y.~Zhang, M.~E. Sherr, C.~Shields, D.~A. Wagner, and W.~Zhou, ``Hidden voice commands,'' in \emph{USENIX Security Symposium}, 2016.

\bibitem{Tu2021ExploringAR}
J.~Tu, H.~Li, X.~Yan, M.~Ren, Y.~Chen, M.~Liang, E.~Bitar, E.~Yumer, and R.~Urtasun, ``Exploring adversarial robustness of multi-sensor perception systems in self driving,'' \emph{ArXiv}, vol. abs/2101.06784, 2021.

\bibitem{Luo2021InteractivePF}
Y.~Luo, M.~Meghjani, Q.~H. Ho, D.~Hsu, and D.~Rus, ``Interactive planning for autonomous urban driving in adversarial scenarios,'' \emph{2021 IEEE International Conference on Robotics and Automation (ICRA)}, pp. 5261--5267, 2021.

\bibitem{Pei2017DeepXploreAW}
K.~Pei, Y.~Cao, J.~Yang, and S.~S. Jana, ``Deepxplore: Automated whitebox testing of deep learning systems,'' \emph{Proceedings of the 26th Symposium on Operating Systems Principles}, 2017.

\bibitem{Zhang2022MachineLT}
J.~Zhang, M.~Harman, L.~Ma, and Y.~Liu, ``Machine learning testing: Survey, landscapes and horizons,'' \emph{IEEE Transactions on Software Engineering}, vol.~48, pp. 1--36, 2022.

\bibitem{Madry2018TowardsDL}
A.~Madry, A.~Makelov, L.~Schmidt, D.~Tsipras, and A.~Vladu, ``Towards deep learning models resistant to adversarial attacks,'' \emph{ICLR}, vol. abs/1706.06083, 2018.

\bibitem{Asyrofi2021CanDT}
M.~H. Asyrofi, Z.~Yang, J.~Shi, C.~W. Quan, and D.~Lo, ``Can differential testing improve automatic speech recognition systems?'' \emph{2021 IEEE International Conference on Software Maintenance and Evolution (ICSME)}, pp. 674--678, 2021.

\bibitem{Gao2020FuzzTB}
X.~Gao, R.~K. Saha, M.~R. Prasad, and A.~Roychoudhury, ``Fuzz testing based data augmentation to improve robustness of deep neural networks,'' \emph{2020 IEEE/ACM 42nd International Conference on Software Engineering (ICSE)}, pp. 1147--1158, 2020.

\bibitem{Ma2018MODEAN}
S.~Ma, Y.~Liu, W.-C. Lee, X.~Zhang, and A.~Y. Grama, ``Mode: automated neural network model debugging via state differential analysis and input selection,'' \emph{Proceedings of the 2018 26th ACM Joint Meeting on European Software Engineering Conference and Symposium on the Foundations of Software Engineering}, 2018.

\bibitem{Tao2020TRADERTD}
G.~Tao, S.~Ma, Y.~Liu, Q.~Xu, and X.~Zhang, ``Trader: Trace divergence analysis and embedding regulation for debugging recurrent neural networks,'' \emph{2020 IEEE/ACM 42nd International Conference on Software Engineering (ICSE)}, pp. 986--998, 2020.

\bibitem{Zhang2017SentimentAA}
L.~Zhang and B.~Liu, ``Sentiment analysis and opinion mining,'' in \emph{Encyclopedia of Machine Learning and Data Mining}, 2017.

\bibitem{Wang2017EmotionRW}
S.~Wang, W.~Wang, J.~Zhao, S.~Chen, Q.~Jin, S.~Zhang, and Y.~Qin, ``Emotion recognition with multimodal features and temporal models,'' \emph{Proceedings of the 19th ACM International Conference on Multimodal Interaction}, 2017.

\bibitem{Bahdanau2015NeuralMT}
D.~Bahdanau, K.~Cho, and Y.~Bengio, ``Neural machine translation by jointly learning to align and translate,'' \emph{ICLR}, vol. abs/1409.0473, 2015.

\bibitem{Wang2022UnderstandingAI}
W.~Wang, W.~Jiao, Y.~Hao, X.~Wang, S.~Shi, Z.~Tu, and M.~R. Lyu, ``Understanding and improving sequence-to-sequence pretraining for neural machine translation,'' in \emph{Annual Meeting of the Association for Computational Linguistics}, 2022.

\bibitem{Jiao2022TencentsMM}
W.~Jiao, Z.~Tu, J.~Li, W.~Wang, J.~tse Huang, and S.~Shi, ``Tencent’s multilingual machine translation system for wmt22 large-scale african languages,'' \emph{WMT}, 2022.

\bibitem{Wang2017TacotronTE}
Y.~Wang, R.~J. Skerry-Ryan, D.~Stanton, Y.~Wu, R.~J. Weiss, N.~Jaitly, Z.~Yang, Y.~Xiao, Z.~Chen, S.~Bengio, Q.~V. Le, Y.~Agiomyrgiannakis, R.~A.~J. Clark, and R.~A. Saurous, ``Tacotron: Towards end-to-end speech synthesis,'' in \emph{Interspeech}, 2017.

\bibitem{Ma2018FPETSFP}
D.~Ma, Z.~Su, W.~Wang, and Y.~Lu, ``Fpets: Fully parallel end-to-end text-to-speech system,'' in \emph{AAAI Conference on Artificial Intelligence}, 2018.

\bibitem{Gupta2020MachineTT}
S.~Gupta, ``Machine translation testing via pathological invariance,'' \emph{2020 IEEE/ACM 42nd International Conference on Software Engineering: Companion Proceedings (ICSE-Companion)}, pp. 107--109, 2020.

\bibitem{He2021TestingMT}
P.~He, C.~Meister, and Z.~Su, ``Testing machine translation via referential transparency,'' \emph{2021 IEEE/ACM 43rd International Conference on Software Engineering (ICSE)}, pp. 410--422, 2021.

\bibitem{Jiao2023IsCA}
W.~Jiao, W.~Wang, J.~tse Huang, X.~Wang, and Z.~Tu, ``Is chatgpt a good translator? a preliminary study,'' \emph{ArXiv}, vol. abs/2301.08745, 2023.

\bibitem{Ribeiro2020BeyondAB}
M.~T. Ribeiro, T.~S. Wu, C.~Guestrin, and S.~Singh, ``Beyond accuracy: Behavioral testing of nlp models with checklist,'' in \emph{ACL}, 2020.

\bibitem{Li2020BERTATTACKAA}
L.~Li, R.~Ma, Q.~Guo, X.~Xue, and X.~Qiu, ``Bert-attack: Adversarial attack against bert using bert,'' \emph{EMNLP}, vol. abs/2004.09984, 2020.

\bibitem{Sun2022ImprovingMT}
Z.~Sun, J.~Zhang, Y.~Xiong, M.~Harman, M.~Papadakis, and L.~Zhang, ``Improving machine translation systems via isotopic replacement,'' \emph{2022 IEEE/ACM 44th International Conference on Software Engineering (ICSE)}, pp. 1181--1192, 2022.

\end{thebibliography}
\end{document}